\documentclass[aps,pra,twocolumn,groupedaddress,showpacs,floatfix]{revtex4-1}

\usepackage{times,amssymb,amsmath}
\usepackage[varg]{txfonts}
\usepackage{amssymb,amsmath}
\usepackage{graphicx}
\usepackage{xspace}
\usepackage{cleveref}
\usepackage{color}
\renewcommand{\vec}[1]{\mathbf{#1}}

\mathchardef\mhyphen="0002D

\renewcommand{\Re}{\text{Re}\,}

\newcommand{\be}{\begin{equation}}
\newcommand{\ee}{\end{equation}}

\newcommand{\w}{\omega}

\newcommand{\diagM}{\raisebox{-10pt}{\includegraphics[width=1.6cm]{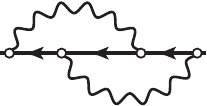}}}
\begin{document}
\title{Vertex corrections for positive-definite spectral functions of 
simple metals}
\author{Y. Pavlyukh}
\affiliation{Institut f\"ur Physik, Martin-Luther-Universit\"at Halle-Wittenberg, 06120 Halle, Germany}
\author{A.-M. Uimonen}
\affiliation{Clarendon Laboratory, Department of Physics, University of Oxford, Parks Road, Oxford, OX1 3PU, UK}
\author{G. Stefanucci}
\affiliation{Dipartimento di Fisica and European Theoretical Spectroscopy Facility (ETSF), 
Universit\`{a} di Roma Tor Vergata,
Via della Ricerca Scientifica 1, 00133 Rome, Italy}
\affiliation{INFN, Sezione di Roma Tor Vergata,
Via della Ricerca Scientifica 1, 00133 Roma, Italy}
\author{R. van Leeuwen}
\affiliation{Department of Physics and European Theoretical Spectroscopy Facility (ETSF),
  Nanoscience Center, University of Jyv{\"a}skyl{\"a},  FI-40014 Jyv{\"a}skyl{\"a}, Finland}
\date{\today}
\begin{abstract}
We present a systematic study of vertex corrections in the homogeneous electron gas at
metallic densities. The vertex diagrams are built using a recently proposed
positive-definite diagrammatic expansion for the spectral function.  The vertex function
not only provides corrections to the well known plasmon and particle-hole scatterings, but
also gives rise to new physical processes such as generation of two plasmon excitations or
the decay of the one-particle state into a two-particles-one-hole state. By an efficient
Monte Carlo momentum integration we are able to show that the additional scattering
channels are responsible for the bandwidth reduction observed in photoemission experiments
on bulk sodium, appearance of the secondary plasmon satellite below the Fermi level, and a
substantial redistribution of spectral weights.  The feasibility of the approach for
first-principles band-structure calculations is also discussed.
\end{abstract}
\maketitle
Starting from the introduction of the notion of quasiparticle
($qp$)~\cite{landau_collected_1967} as an elemental excitation in Fermi
liquids~\cite{nozieres_theory_1999} we have almost a complete picture of its on-shell
properties~\cite{giuliani_quantum_2005,stefanucci_nonequilibrium_2013}.  Quantum Monte
Carlo simulations~\cite{moroni_static_1995,holzmann_momentum_2011}, phase
diagrams~\cite{wigner_interaction_1934,overhauser_new_1959,ortiz_zero_1999,
  trail_unrestricted_2003,baguet_hartree-fock_2013,zhang_hartree-fock_2008}, structure
factors~\cite{gori-giorgi_analytic_2000}, and effective interparticle
interactions~\cite{gori-giorgi_short-range_2001} contributed to our knowledge of Fermi
liquids and to the development of the density
functionals~\cite{petersilka_excitation_1996,onida_electronic_2002}.

Despite these surpluses we still have a poor knowledge of the energy- and
momentum-resolved spectral function $A(k,\omega)$ away from the on-shell manifold, i.\,e.,
when $\omega\not\approx \varepsilon_k$. In angular resolved photoemission this is the
regime where electrons with reduced energy (as compared to the prediction based on
band-structure and energy balance) are observed. Self-consistent ($sc$) perturbation
theory, e.g., $sc$-$GW$~\cite{hedin_new_1965}, accurately predicts total
energies~\cite{almbladh_variational_1999,garcia-gonzalez_self-consistent_2001,dahlen_self-consistent_2005}
and it is fully conserving at the one-particle level, a crucial property in the
description of transport phenomena~\cite{stan_levels_2009}. However, for spectral
properties $sc$ schemes do not show the expected improvement over simpler one-shot
calculations~\cite{lundqvist_single-particle_1967,lundqvist_single-particle_1967-1,lundqvist_single-particle_1968}.
In fact, they suffer from serious drawbacks: the incoherent background in the spectral
function gains weight at the expenses of the $qp$ peak~\cite{holm_fully_1998}, the $qp$
energy does not agree with experiments (overestimating the bandwidth of simple
metals)~\cite{yasuhara_why_1999,takada_inclusion_2001}, and the screened interaction does
not obey the $f$-sum rule~\cite{kwong_real-time_2000,pal_conserving_2009}. It was then
proposed that self-energy (SE) diagrams with vertex corrections may cancel the spurious
$sc$
effects~\cite{mahan_electron-electron_1989,mahan_gw_1994,bobbert_lowest-order_1994,holm_fully_1998,schindlmayr_systematic_1998}.
This fueled a number of notable attempts to include the vertex function in a \emph{model}
fashion: using the plasmon model for the screened
interaction~\cite{minnhagen_aspects_1975}, neglecting the incoherent part of the electron
spectral function~\cite{shirley_self-consistent_1996}, employing the Ward identities and a
model form of the exchange-correlation
kernel~\cite{takada_inclusion_2001,takada_dynamical_2002,bruneval_many-body_2005,maebashi_analysis_2011},
or the \emph{sc} cumulant
expansion~\cite{holm_self-consistent_1997,kas_cumulant_2014,caruso_cumulant_2016}.
Although these methods clarified a number of issues, they did not provide an exhaustive
picture~\footnote{For instance the cumulant expansion is exact for deep core states
  interacting with plasmons and leads to the spectrum with equally spaced
  satellites~\cite{langreth_singularities_1970}. Yet, this assumption is less justified
  for the valence band excitations overestimating the weight of higher order plasmon
  satellites (something that can be partially cured by taking multiple plasmon branches
  and their dispersion into account~\cite{cini_coherent_1986,guzzo_multiple_2014}).}.  The
major obstacle for a full-fledged vertex calculation, besides numerical complexity, is the
issue of \emph{negative spectral densities}, first noted by
Minnhagen~\cite{minnhagen_vertex_1974,minnhagen_aspects_1975} and only recently solved by
us using a positive-definite diagrammatic
expansion~\cite{PSDtot,uimonen_diagrammatic_2015}.  Our solution merges many-body
perturbation theory (MBPT) and scattering theory, thus returning a positive-semidefinite
(PSD) spectral function by construction.

With the PSD tool at our disposal, in this Letter we investigate the influence of vertex
corrections on the spectral function of the homogeneous electron gas (HEG), paving the way
towards first-principles correlated calculations of band-structures. We demonstrate that
the vertex function leads to a number of novel physical phenomena which cannot be reduced
to mere self-consistency cancellations. Stochastic methods, long been used to describe
integral properties, are shown to be well suited for the calculation of spectral features
too.  In fact, our Monte Carlo momentum integration is so efficient that the numerical
part of the calculations does not pose any difficulty.

\begin{figure}[]
\centering
\includegraphics[width=\columnwidth]{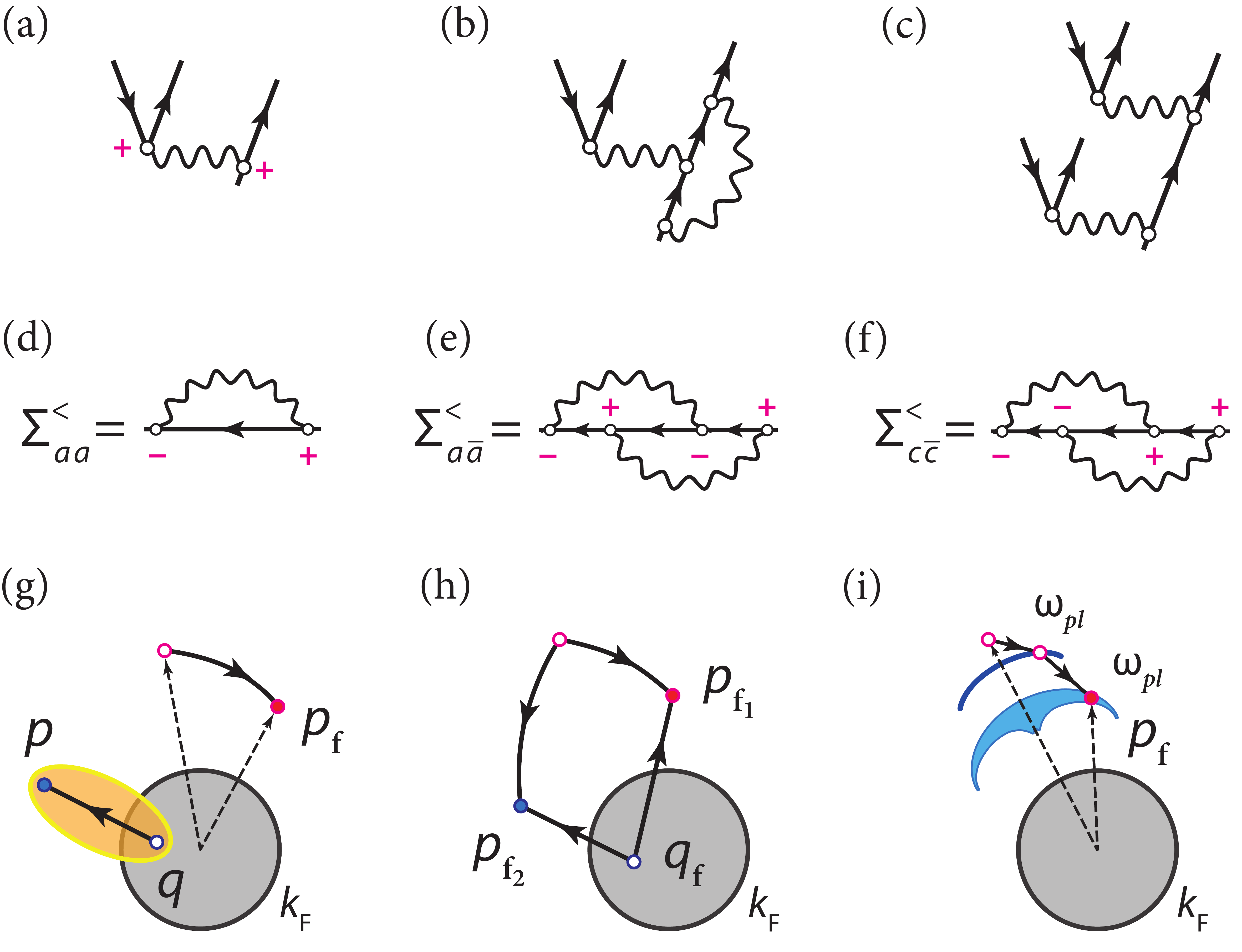}
\caption{\label{fig:diag}(Color online) (a-c) The half-diagrams emerging from the
  bisection of the $\Sigma^{(2)}$ partitions (wiggly lines denote the screened
  interaction).  (d-f) Three partitions of the PSD self-energy and (g-i) their momentum
  space representation.}
\end{figure}

Let us motivate and discuss the PSD approximation used in this work. In terms of {\em
  dressed} electronic propagators and {\em screened} interaction $W$ there is a single
second order SE diagram $\Sigma^{(2)}=$\,\diagM.  Its straightforward inclusion, however,
yields negative spectra in some frequency regions. This prohibits the usual probability
interpretation and, even worse, it jeopardizes \emph{sc} calculations since the resulting
Green's function (GF) has the wrong analytic structure~\cite{KvL.2016}. The key idea of
the PSD scheme~\cite{PSDtot,uimonen_diagrammatic_2015} consists in (1) writing a SE
diagram as the sum of its {\em partitions}, i.e., diagrams with particle and hole
propagators, (2) bisect each partition into two half-diagrams, (3) add the missing
half-diagrams to form a perfect square, and (4) glue the half-diagrams back.  For
$\Sigma^{(2)}$ the half-diagrams, see Fig.~\ref{fig:diag}(a-c), contain scatterings with
up to three particles and two holes in the final state~\cite{PSDtot}.  The SE partitions
stem from the interference between these scatterings and after the PSD treatment one
obtains partitions up to the {\em fourth} order in $W$, see Ref.~\onlinecite{PSDtot} for
the full list.  Among them there are three which deserve special attention.  $\Sigma_{aa}$
in Fig.~\ref{fig:diag}(d) results from the interference of scattering (a) with itself. As
illustrated in Fig.~\ref{fig:diag}(g) $\Sigma_{aa}$ involves a particle-hole ($ph$) pair
(orange area) or a plasmon in the final state (this is the first order effect described by
the $GW$ SE).  The plus and minus vertices in the SE partitions have the purpose of
distinguishing the constituent half-diagrams (resulting from the cut of all propagators
with $+/-$ and $-/+$ vertices).  $\Sigma_{a\bar{a}}$ in Fig.~\ref{fig:diag}(e) is formed
by the interference between the scattering (a), leading to two-particle-one-hole
($p_{f_1}$-$p_{f_2}$-$q_f$) final state, and the same scattering with interchanged
particle momenta (indicated with $\bar{a}$), see Fig.~\ref{fig:diag}(h).  Finally,
$\Sigma_{c\bar{c}}$ in Fig.~\ref{fig:diag}(f) is formed by the interference between the
scattering (c), in which a particle loses its energy by exciting 2 $ph$ pairs, 2 plasmons
or a mixture of them, and the same scattering with intechanged particle and hole momenta,
see Fig.~\ref{fig:diag}(i).  Plasmon generation is a dominant second order scattering
process although it has a severely limited phase-space (dark blue line and light-blue
area) due to energy and momentum conservation.  Higher order terms in $W$ (up to fourth
order) arise from other interferences and are needed to assure the overall
positivity~\cite{PSDtot}.  In general the PSD procedure leads to a manifestly positive
Fermi Golden rule form of the SE, $\Sigma^{<}(k,\omega)\sim\sum_{n,f}
\Gamma^{(n)}(k,\omega)|1+
r_s\gamma^{(n)}_1+r_s^2\gamma^{(n)}_2+\ldots|^2\delta(\omega+\epsilon_k-E_{f}^{(n)})$,
where the sum runs over all final states of energy $E_{f}^{(n)}$ with $(n+1)$-particles
and $n$-holes ($r_{s}$ being Wigner-Seitz radius).  The role of high order diagrams is
two-fold: they bring new scattering mechanisms into play (hence new rates $\Gamma^{(n)}$)
and renormalize them through the perturbative corrections $\gamma^{(n)}_i$.

We already mentioned that one of the motivations for including diagrams beyond $GW$ is the
excessive broadening of the spectral features when the level of self-consistency is
increased, e.g., $G^{(0)}W^{(0)}\to GW^{(0)} \to GW$.  As a full $sc$ calculation of the
PSD SE of Ref.~\cite{PSDtot} is out of reach, we partially account for self-consistency by
using a $G^{(0)}W^{(0)}$ GF (finite $qp$-broadening and plasmon satellites) and an RPA
screened interaction.  Our calculations indicate that higher-order diagrams (aside from
bringing in new spectral features) counteract the undesired $sc$ effects, thus suggesting
the occurrence of sizable cancellations.
\begin{figure}[b!]
\centering
\includegraphics[width=\columnwidth]{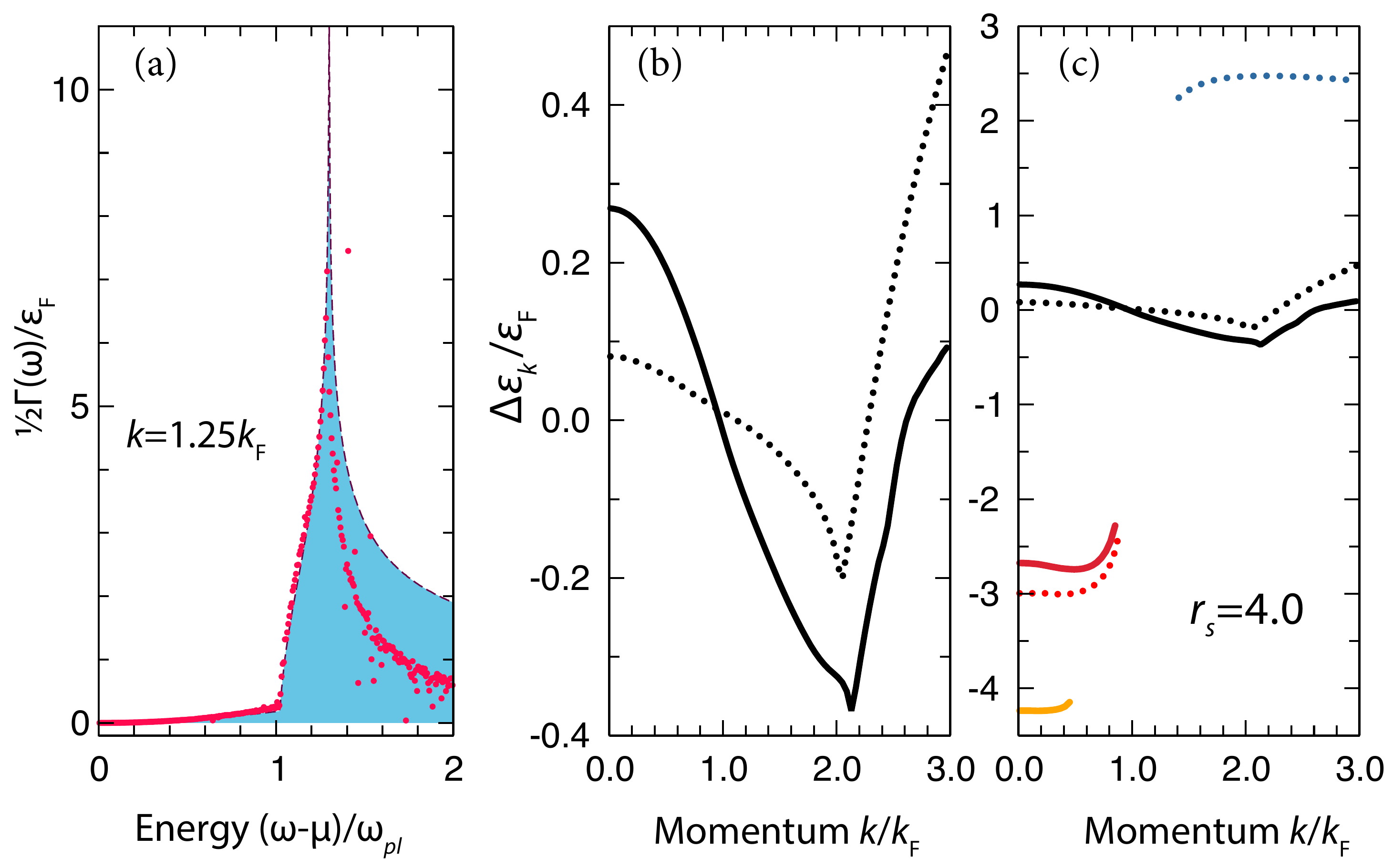}
\caption{\label{fig:Ek} (Color online) (a) Rate $\Gamma(k,\w)$ for $k=1.25k_{F}$
  calculated from the PSD SE of Ref.~\onlinecite{PSDtot} with $G^{(0)}W^{(0)}$ GF (red
  dots) and from the SE $\Sigma=\Sigma_{a\bar{a}}+\Sigma_{c\bar{c}}+\Sigma_{aa}$ with $qp$
  GF (dashed). (b) $qp$ energy correction
  $\Delta\epsilon_{k}=\epsilon_{k}-\epsilon_{k}^{(0)}$ and (c) plasmon dispersions for
  $G^{(0)}W^{(0)}$ (dotted) and our vertex approximation (full). The corrections to $\mu$
  (with respect to the mean-field value) are $\Delta\mu=-1.76\epsilon_F$ and $\Delta
  \mu=-1.81\epsilon_F$ respectively. }
\end{figure}
We then explore the possibility of producing the PSD results with {\em less} diagrams and
{\em bare} GF's. In the bare GF the chemical potential $\mu$ is iteratively adjusted by
imposing that the energy of states on the Fermi sphere (where the discontinuity in the
momentum distribution $n_k$ occurs) is exactly equal to $\mu$~\footnote{We start with
  zeroth approximation $\Delta\mu^{(0)}=\Re\Sigma(k_F,1/2k_F^2)$ and perform two more
  calculations for $k=k_F\pm\Delta k$, where $\Delta k$ is a small number, typically a few
  percents of the Fermi momentum. The refined chemical potential shift is then given by
  $\Delta\mu=\Delta\mu^{(0)}+\frac12(\Delta\epsilon_{k_F+\Delta
    k}+\Delta\epsilon_{k_F-\Delta k})$, where $\Delta\epsilon_{k}$ is the correlational
  shift.}.  In Fig.~\ref{fig:Ek}(a) we compare the rate
\begin{equation}
\Gamma(k,\omega)\equiv i[\Sigma^{\rm R}(k,\omega)-\Sigma^{\rm A}(k,\omega)],
\label{gamma}
\end{equation}
as obtained from the PSD diagrams of Ref.~\cite{PSDtot} with $G^{(0)}W^{(0)}$ GF and from
the much simpler $\Sigma=\Sigma_{a\bar{a}}+\Sigma_{c\bar{c}}+\Sigma_{aa}$ with $qp$ GF (in
both cases we used an RPA $W$).  The left flank and the hight of the peak are in perfect
agreement.  At energies in the region of plasmon satellites the full PSD rate decays
faster but the trend is similar and the impact of this discrepancy on the spectral
function is only minor. More calculations at different $k$ (not shown) confirm the
agreement between the two SEs.  We therefore infer that the relevant scattering mechanisms
for a positive-conserving, leading-order vertex correction are those of
Fig.~\ref{fig:diag}(g-i). This reduction of diagrams represents an important advance in
view of correlated band-structure calculations of solids.  In the following we use the
vertex correction of Fig.~\ref{fig:diag}(d-f) to calculate $qp$ and plasmon energy
dispersions, spectral function, scattering rate, renormalization factor and momentum
distribution.

\paragraph{Results}

\begin{figure}[] 
\centering
\includegraphics[width=0.9\columnwidth]{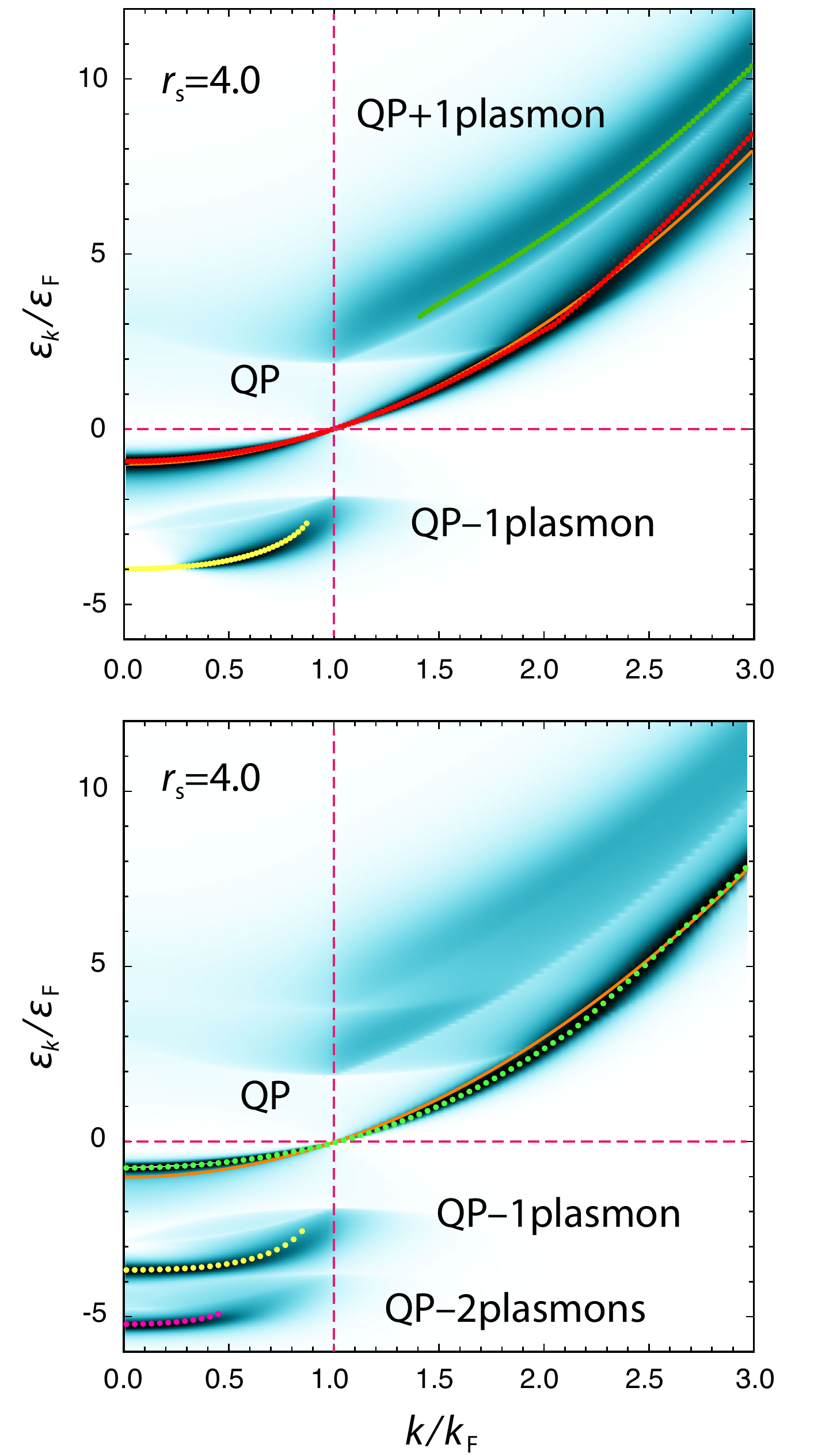}
\caption{\label{fig:Akw} (Color online) Energy- and momentum-resolved spectral function
  without (top) and with (bottom) vertex corrections. Solid lines denote the free electron
  dispersion. Dots denote the solutions of the real Dyson equation, see main text. For
  some momentum values multiple solutions (marked with different colors) are obtained.}
\end{figure}

The electron density and the dimensionality completely determine the properties of the
HEG; in the 3d case they fix the Fermi momentum and energy as follows: $k_F=(\alpha
r_s)^{-1}$ and $\epsilon_F=1/2(\alpha r_s)^{-2}$ with $\alpha=[4/(9\pi)]^{1/3}$.  We
consider the case of metallic densities $r_s=4.0\,a_B$ appropriate for, e.\,g., bulk Na
metal. Angle-resolved photoemission experiments have pointed out a substantial narrowing
of the occupied band in sodium~\cite{lyo_quasiparticle_1988}. In
Ref.~\cite{yasuhara_why_1999} the origin of this narrowing was ascribed to the effects of
electron correlations on the unoccupied bands.  Such conclusion is by no means obvious as
fully $sc$-$GW$ results~\cite{holm_fully_1998} indicate the opposite trend (bandwidth
larger by 20\% as compared to the noninteracting electron dispersion).  As it was pointed
out in Ref.~\cite{takada_inclusion_2001}, vertex corrections rather than screening are
crucial to reproduce the experimentally observed dispersion. Our diagrammatic
approximation confirms this fact, providing a bandwidth reduction of 27.5\% (as compared
to the noninteracting case), see Fig.~\ref{fig:Ek}(b). Fig.~\ref{fig:Ek}(c) also shows the
dispersion of plasmon satellites (red and blue curves).  In the calculations with vertex
corrections (solid) the high-energy plasmon branch smears out and, unlike in the
$G^{(0)}W^{(0)}$ approximation (dotted), there is no real solution to the Dyson
equation. We further observe an upward renormalization of the $G^{(0)}W^{(0)}$ low-energy
plasmon branch, in agreement with experiment~\cite{aryasetiawan_multiple_1996}, as well as
the emergence a second branch at lower energy (orange).

Vertex corrections have a sizable impact on the energy and momentum resolved spectral
function $A(k,\omega)=i[G^\mathrm{R}-G^\mathrm{A}](k,\omega)$.  In Fig.~\ref{fig:Akw} we
display the color plot of $A(k,\omega)$ without (top) and with (bottom) vertex
corrections.  The dotted lines denote the solutions of the real part of the Dyson equation
(used to produce the curves in Fig.~\ref{fig:Ek}):
$\omega+\Delta\mu-\epsilon_k=\Re[\Sigma^\mathrm{R}(k,\omega)]$.  Appearance of the 2nd
plasmon satellite \emph{below} $\mu$, redistribution of the spectral weight between 1st
and 2nd satellite, and further broadening of plasmonic spectral features \emph{above}
$\mu$ are the most important findings of this work. They confirm the plasmon-pole model
analysis of Ref.~\cite{shirley_self-consistent_1996} that predicted only hole satellites
and much broader particle features.  Our results call into question the cumulant
parameterization of the spectral function in Ref.~\cite{holm_self-consistent_1997} where
no distinction between hole and particle features is made.

\begin{figure*}[] 
\centering
\includegraphics[width=\textwidth]{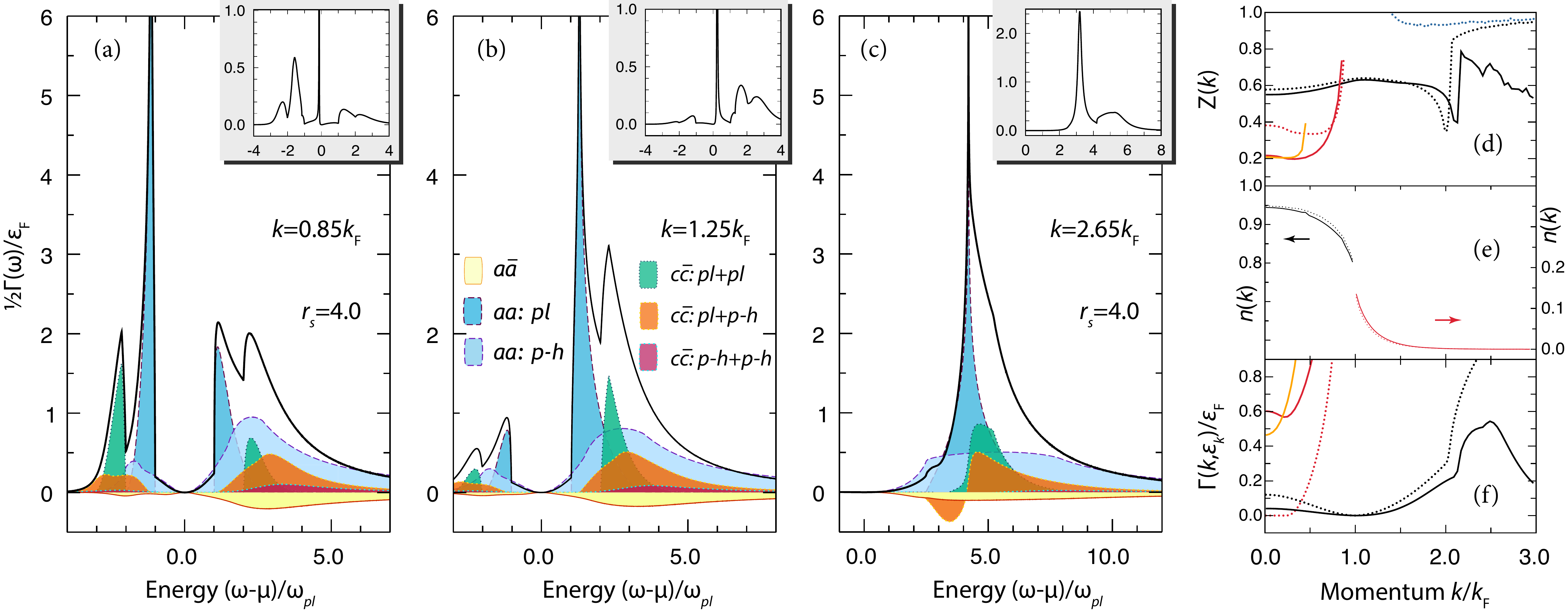}
\caption{\label{fig:sgm} (Color online) (a-c) $\Gamma(k,\omega)$ for 3 momentum
  values. Black line is the total contribution from SE diagrams $a\bar{a}$, $c\bar{c}$,
  and $aa$. The contribution of each diagram is separated according to the intermediate
  states ($ph$ or plasmons) involved, and it is displayed in different colors.  (d)
  Renormalization factor of the $qp$ and $pl$ excitations.  (e) Momentum occupation number
  $n_k$.  (f) Broadening of $qp$ and $pl$ excitations. For panels (d-f) we show results
  using $G^{(0)}W^{(0)}$ (dotted) and our vertex approximation (full).}
\end{figure*}

It is interesting to notice that vertex corrections make the $qp$-peak sharper.  This can
be inferred from the explicit SE expression or from the rate $\Gamma$ of Eq.~(\ref{gamma})
which we plot in Fig.~\ref{fig:sgm}(a-c) for three different values of the
momentum. Plasmons do not contribute to the on-shell properties at energies around $\mu$
because they carry finite energy at zero momentum ($\omega_{pl}(q=0)=1.881\epsilon_F$ for
$r_s=4$). Instead, the life-time of $qp$s in the vicinity of the Fermi sphere is mainly
determined by $\Sigma_{aa}$ ($GW$ SE) involving $ph$ production or by $\Sigma_{a\bar{a}}$
($qp\rightarrow 3 qp$).  The latter, shown as yellow shaded curve in
Fig.~\ref{fig:sgm}(a-c), contributes with negative sign and leads to the observed
reduction of $\Gamma$ (hence an enhancement of the $qp$ peak)~\footnote{If $W$ in
  $\Sigma_{a\bar{a}}$ is replaced by the bare Coulomb interaction the so-called
  second-order exchange SE is obtained. Its on-shell value is density independent and it
  is known
  analytically~\cite{onsager_integrals_1966,ziesche_self-energy_2007,glasser_analysis_2007},
  $\Sigma_{2x}(k_F,1/2k_F^2)=\big[2\pi^2\ln(2)/3-3\zeta(3)\big]/4\pi^2$. This result
  represents a useful check for our numerical algorithms.}.  Such a behavior (alternating
series in $\alpha r_s$) is typical of many perturbation theories. Notice that
$\Sigma_{aa}+\Sigma_{a\bar{a}}$ also dominates the asymptotic ($\w\to\infty$)
behavior. The scattering with generation of 2 plasmons, contained in $\Sigma_{c\bar{c}}$,
plays a crucial role for the off-shell properties as it gives rise to new spectral peaks,
see green shaded curve.

In the vicinity of a $qp$ or plasmon ($pl$) peak the spectral function acquires the
form~\cite{pavlyukh_initial_2013}:
\begin{align}
A(k,\omega)&=Z^{(\alpha)}(k)\frac{1/\tau^{(\alpha)}_k}
{(\omega-\omega^{(\alpha)}_k)^2+1/(2\tau_k^{(\alpha)})^2},
\end{align}
where $\alpha=qp,pl$ and $\omega^{(qp)}_{k}=\epsilon_{k}$ whereas
$\omega^{(pl)}_{k}=\omega_{pl}(k)$ is the dispersion of plasmon satellites.  This
expression contains two quantities of physical interest that we computed using our vertex
function: the renormalization factor
\begin{align}
Z^{(\alpha)}(k)=\left(1-\frac{\partial}{\partial\omega}
\left.\Re\Sigma^\mathrm{R}(k,\omega)\right|_{\omega=\omega^{(\alpha)}_k}\right)^{-1},
\label{eq:Zk}
\end{align}
and the broadening of the $qp$ or $pl$ excitations
$1/\tau^{(\alpha)}_k=Z^{(\alpha)}(k)\Gamma(k,\omega^{(\alpha)}_k)$.

The renormalization factor is shown in Fig.~~\ref{fig:sgm}(d).  At the band bottom ($k=0$)
$G^{(0)}W^{(0)}$ gives only one plasmon satellite $Z^{(pl)}=0.382$ whereas our vertex
approximation gives two satellites with comparable weight $Z^{(pl)}=0.217$ and
$Z^{(2pl)}=0.207$.  Furthermore, the $qp$ weight is reduced from $Z^{(qp)}=0.578$ (in
$G^{(0)}W^{(0)}$) to $Z^{(qp)}=0.550$ indicating that the incoherent part of the spectrum
gains weight.  These two effects cannot be seen in the cumulant expansion
scheme~\cite{holm_self-consistent_1997} which suppresses the $Z$ of higher plasmon
satellites according to the Poissonian distribution~\cite{aryasetiawan_multiple_1996} and,
due to the neglect of the coupling between particle and hole seas, yields the same
$Z^{(qp)}$ as in $G^{(0)}W^{(0)}$.

For $k=k_{F}$ vertex corrections reduce only slightly the $G^{(0)}W^{(0)}$ $qp$
renormalization factor.  It is known that $sc$-$GW$ overestimates ($Z^{(qp)}=0.793$) the
already good $G^{(0)}W^{(0)}$ value $Z^{(qp)}=0.638$ (our calculation) or $Z^{(qp)}=0.646$
(Hedin~\cite{hedin_new_1965}). The proposed approximation to the vertex gives
$Z^{(qp)}=0.628$, which thus remains rather close to QMC results $0.64$ to $0.69$ (at
$r_s=3.99\,a_B$)~\cite{holzmann_momentum_2011}. At the Fermi momentum $Z^{(qp)}$ can also
be deduced from the discontinuity of the momentum distribution function
$n_k$~\cite{mahaux_theoretical_1992,gori_giorgi_momentum_2002,olevano_momentum_2012}.  In
Fig.~\ref{fig:sgm}(e) we show $n_{k}$ as obtained by a straightforward integration of the
smooth part of the spectral function,
$n_k=\int_{-\infty}^{\mu}\frac{d\omega}{2\pi}A(k,\omega)$, and by adding the singular
contributions analytically. The $G^{(0)}W^{(0)}$ and vertex results are almost
indistinguishable.

We finally analyze in Fig.~\ref{fig:sgm}(f) the quasiparticle life-time, a measure of
electronic correlations~\cite{echenique_theory_2000,qian_lifetime_2005}. In \emph{ab
  initio} calculations for realistic systems this quantity is typically estimated using
the $G_0W_0$
approximation~\cite{zhukov_lifetimes_2002,pavlyukh_decay_2008,pavlyukh_communication:_2011}.
However, there have also been attempts to go beyond this level of theory by, e.g.,
including $T$-matrix diagrams. In Ref.~\cite{nechaev_variational_2005} a \emph{reduction}
of $qp$ life-time (increase of $\Gamma$ by 50\% (70\%) in relation to $GW$ for $r_s =2.07
(4.86)$) has been predicted and explained by "the multiple scattering".  Our findings show
the opposite trend, i.e., an increase of the $qp$ life-time (reduction of $\Gamma$ by
$-$50\% in relation to $GW$ for $r_s=4$).

\paragraph{Conclusions}
Numerous authors emphasized that the inclusion of the vertex function should remedy the
drawbacks of self-consistent
calculations~\cite{holm_fully_1998,mahan_electron-electron_1989,verdozzi_evaluation_1995,
  takada_inclusion_2001}. Using our recently proposed diagrammatic analysis we have been
able to confirm these expectations and show that this is only a part of the whole
picture. Additionally, other second-order processes appear. They can be best described in
the language of scattering theory with the link provided by the PSD
formalism~\cite{PSDtot,uimonen_diagrammatic_2015}.

We fully characterized the spectral function of 3d HEG in the $k-\omega$ plane. The main
original features that we found are: a second plasmon satellite for holes, redistribution
of the spectral weight between hole satellites, reduction of the plasmon spectral weight
for particles, bandwidth reduction of the main $qp$-band. So far these effects have only
been partially captured by other, non-diagrammatic methods.  Our proposed approach has a
universal character and can be extended to first-principle calculations of metals.  In
fact, in going from continuous to discrete translational symmetry the functions simply
turn into matrix functions (e.\,g., $\Sigma(\vec k,\omega)\to \Sigma_{GG'}(\vec
k,\omega)$), something which does not pose any conceptual difficulties for Monte Carlo
momentum integration~\footnote{Due to the long-range character of the Coulomb interaction
  (bare and to a lesser extent screened one) the integrations need to be extended beyond
  the boundaries of the first Brillouin zone. For $GW$ calculations the fast Fourier
  method have been proposed~\cite{rojas_space-time_1995}. For higher dimensional
  integrals, such as in the present calculations, this approach becomes impractical, but
  can be remedied by the Monte Carlo integration featuring excellent scalability.}.
Alkali metals for which a vertex function was partially included (typically using a model
exchange-correlation kernel~\cite{northrup_theory_1987,lischner_effect_2014}) is a logical
next step for our method.

\subsection*{Acknowledgements}

We acknowledge CSC - IT Center for Science, Finland, for computational resources.
Y.P. acknowledges support by the DFG through grant No. PA 1698/1-1.  G.S. acknowledges
funding by MIUR FIRB Grant No. RBFR12SW0J and EC funding through the RISE Co-ExAN
(GA644076).  R.vL. would like to thank the Academy of Finland for support.


\begin{thebibliography}{71}%
\makeatletter
\providecommand \@ifxundefined [1]{%
 \@ifx{#1\undefined}
}%
\providecommand \@ifnum [1]{%
 \ifnum #1\expandafter \@firstoftwo
 \else \expandafter \@secondoftwo
 \fi
}%
\providecommand \@ifx [1]{%
 \ifx #1\expandafter \@firstoftwo
 \else \expandafter \@secondoftwo
 \fi
}%
\providecommand \natexlab [1]{#1}%
\providecommand \enquote  [1]{``#1''}%
\providecommand \bibnamefont  [1]{#1}%
\providecommand \bibfnamefont [1]{#1}%
\providecommand \citenamefont [1]{#1}%
\providecommand \href@noop [0]{\@secondoftwo}%
\providecommand \href [0]{\begingroup \@sanitize@url \@href}%
\providecommand \@href[1]{\@@startlink{#1}\@@href}%
\providecommand \@@href[1]{\endgroup#1\@@endlink}%
\providecommand \@sanitize@url [0]{\catcode `\\12\catcode `\$12\catcode
  `\&12\catcode `\#12\catcode `\^12\catcode `\_12\catcode `\%12\relax}%
\providecommand \@@startlink[1]{}%
\providecommand \@@endlink[0]{}%
\providecommand \url  [0]{\begingroup\@sanitize@url \@url }%
\providecommand \@url [1]{\endgroup\@href {#1}{\urlprefix }}%
\providecommand \urlprefix  [0]{URL }%
\providecommand \Eprint [0]{\href }%
\providecommand \doibase [0]{http://dx.doi.org/}%
\providecommand \selectlanguage [0]{\@gobble}%
\providecommand \bibinfo  [0]{\@secondoftwo}%
\providecommand \bibfield  [0]{\@secondoftwo}%
\providecommand \translation [1]{[#1]}%
\providecommand \BibitemOpen [0]{}%
\providecommand \bibitemStop [0]{}%
\providecommand \bibitemNoStop [0]{.\EOS\space}%
\providecommand \EOS [0]{\spacefactor3000\relax}%
\providecommand \BibitemShut  [1]{\csname bibitem#1\endcsname}%
\let\auto@bib@innerbib\@empty
\bibitem [{\citenamefont {Landau}\ and\ \citenamefont
  {Haar}(1967)}]{landau_collected_1967}%
  \BibitemOpen
  \bibfield  {author} {\bibinfo {author} {\bibfnamefont {L.~D.}\ \bibnamefont
  {Landau}}\ and\ \bibinfo {author} {\bibfnamefont {D.~t.}\ \bibnamefont
  {Haar}},\ }\href@noop {} {\emph {\bibinfo {title} {Collected papers of {L}.
  {D}. {Landau}}}}\ (\bibinfo  {publisher} {Gordon and Breach},\ \bibinfo
  {address} {New York; London; Paris},\ \bibinfo {year} {1967})\BibitemShut
  {NoStop}%
\bibitem [{\citenamefont {Nozi\`{e}res}\ and\ \citenamefont
  {Pines}(1999)}]{nozieres_theory_1999}%
  \BibitemOpen
  \bibfield  {author} {\bibinfo {author} {\bibfnamefont {P.}~\bibnamefont
  {Nozi\`{e}res}}\ and\ \bibinfo {author} {\bibfnamefont {D.}~\bibnamefont
  {Pines}},\ }\href@noop {} {\emph {\bibinfo {title} {The theory of quantum
  liquids}}},\ Advanced book classics\ (\bibinfo  {publisher} {Westview Press,
  Perseus Books Group},\ \bibinfo {address} {Boulder, CO},\ \bibinfo {year}
  {1999})\BibitemShut {NoStop}%
\bibitem [{\citenamefont {Giuliani}\ and\ \citenamefont
  {Vignale}(2005)}]{giuliani_quantum_2005}%
  \BibitemOpen
  \bibfield  {author} {\bibinfo {author} {\bibfnamefont {G.}~\bibnamefont
  {Giuliani}}\ and\ \bibinfo {author} {\bibfnamefont {G.}~\bibnamefont
  {Vignale}},\ }\href@noop {} {\emph {\bibinfo {title} {Quantum theory of the
  electron liquid}}}\ (\bibinfo  {publisher} {Cambridge University Press},\
  \bibinfo {address} {Cambridge, UK},\ \bibinfo {year} {2005})\BibitemShut
  {NoStop}%
\bibitem [{\citenamefont {Stefanucci}\ and\ \citenamefont {van
  Leeuwen}(2013)}]{stefanucci_nonequilibrium_2013}%
  \BibitemOpen
  \bibfield  {author} {\bibinfo {author} {\bibfnamefont {G.}~\bibnamefont
  {Stefanucci}}\ and\ \bibinfo {author} {\bibfnamefont {R.}~\bibnamefont {van
  Leeuwen}},\ }\href@noop {} {\emph {\bibinfo {title} {Nonequilibrium
  {Many}-{Body} {Theory} of {Quantum} {Systems}: {A} {Modern}
  {Introduction}}}}\ (\bibinfo  {publisher} {Cambridge University Press},\
  \bibinfo {address} {Cambridge},\ \bibinfo {year} {2013})\BibitemShut
  {NoStop}%
\bibitem [{\citenamefont {Moroni}\ \emph {et~al.}(1995)\citenamefont {Moroni},
  \citenamefont {Ceperley},\ and\ \citenamefont
  {Senatore}}]{moroni_static_1995}%
  \BibitemOpen
  \bibfield  {author} {\bibinfo {author} {\bibfnamefont {S.}~\bibnamefont
  {Moroni}}, \bibinfo {author} {\bibfnamefont {D.~M.}\ \bibnamefont
  {Ceperley}}, \ and\ \bibinfo {author} {\bibfnamefont {G.}~\bibnamefont
  {Senatore}},\ }\href@noop {} {\bibfield  {journal} {\bibinfo  {journal}
  {Phys. Rev. Lett.}\ }\textbf {\bibinfo {volume} {75}},\ \bibinfo {pages}
  {689} (\bibinfo {year} {1995})}\BibitemShut {NoStop}%
\bibitem [{\citenamefont {Holzmann}\ \emph {et~al.}(2011)\citenamefont
  {Holzmann}, \citenamefont {Bernu}, \citenamefont {Pierleoni}, \citenamefont
  {McMinis}, \citenamefont {Ceperley}, \citenamefont {Olevano},\ and\
  \citenamefont {Delle~Site}}]{holzmann_momentum_2011}%
  \BibitemOpen
  \bibfield  {author} {\bibinfo {author} {\bibfnamefont {M.}~\bibnamefont
  {Holzmann}}, \bibinfo {author} {\bibfnamefont {B.}~\bibnamefont {Bernu}},
  \bibinfo {author} {\bibfnamefont {C.}~\bibnamefont {Pierleoni}}, \bibinfo
  {author} {\bibfnamefont {J.}~\bibnamefont {McMinis}}, \bibinfo {author}
  {\bibfnamefont {D.~M.}\ \bibnamefont {Ceperley}}, \bibinfo {author}
  {\bibfnamefont {V.}~\bibnamefont {Olevano}}, \ and\ \bibinfo {author}
  {\bibfnamefont {L.}~\bibnamefont {Delle~Site}},\ }\href@noop {} {\bibfield
  {journal} {\bibinfo  {journal} {Phys. Rev. Lett.}\ }\textbf {\bibinfo
  {volume} {107}},\ \bibinfo {pages} {110402} (\bibinfo {year}
  {2011})}\BibitemShut {NoStop}%
\bibitem [{\citenamefont {Wigner}(1934)}]{wigner_interaction_1934}%
  \BibitemOpen
  \bibfield  {author} {\bibinfo {author} {\bibfnamefont {E.}~\bibnamefont
  {Wigner}},\ }\href@noop {} {\bibfield  {journal} {\bibinfo  {journal} {Phys.
  Rev.}\ }\textbf {\bibinfo {volume} {46}},\ \bibinfo {pages} {1002} (\bibinfo
  {year} {1934})}\BibitemShut {NoStop}%
\bibitem [{\citenamefont {Overhauser}(1959)}]{overhauser_new_1959}%
  \BibitemOpen
  \bibfield  {author} {\bibinfo {author} {\bibfnamefont {A.~W.}\ \bibnamefont
  {Overhauser}},\ }\href@noop {} {\bibfield  {journal} {\bibinfo  {journal}
  {Phys. Rev. Lett.}\ }\textbf {\bibinfo {volume} {3}},\ \bibinfo {pages} {414}
  (\bibinfo {year} {1959})}\BibitemShut {NoStop}%
\bibitem [{\citenamefont {Ortiz}\ \emph {et~al.}(1999)\citenamefont {Ortiz},
  \citenamefont {Harris},\ and\ \citenamefont {Ballone}}]{ortiz_zero_1999}%
  \BibitemOpen
  \bibfield  {author} {\bibinfo {author} {\bibfnamefont {G.}~\bibnamefont
  {Ortiz}}, \bibinfo {author} {\bibfnamefont {M.}~\bibnamefont {Harris}}, \
  and\ \bibinfo {author} {\bibfnamefont {P.}~\bibnamefont {Ballone}},\
  }\href@noop {} {\bibfield  {journal} {\bibinfo  {journal} {Phys. Rev. Lett.}\
  }\textbf {\bibinfo {volume} {82}},\ \bibinfo {pages} {5317} (\bibinfo {year}
  {1999})}\BibitemShut {NoStop}%
\bibitem [{\citenamefont {Trail}\ \emph {et~al.}(2003)\citenamefont {Trail},
  \citenamefont {Towler},\ and\ \citenamefont
  {Needs}}]{trail_unrestricted_2003}%
  \BibitemOpen
  \bibfield  {author} {\bibinfo {author} {\bibfnamefont {J.~R.}\ \bibnamefont
  {Trail}}, \bibinfo {author} {\bibfnamefont {M.~D.}\ \bibnamefont {Towler}}, \
  and\ \bibinfo {author} {\bibfnamefont {R.~J.}\ \bibnamefont {Needs}},\
  }\href@noop {} {\bibfield  {journal} {\bibinfo  {journal} {Phys. Rev. B}\
  }\textbf {\bibinfo {volume} {68}},\ \bibinfo {pages} {045107} (\bibinfo
  {year} {2003})}\BibitemShut {NoStop}%
\bibitem [{\citenamefont {Baguet}\ \emph {et~al.}(2013)\citenamefont {Baguet},
  \citenamefont {Delyon}, \citenamefont {Bernu},\ and\ \citenamefont
  {Holzmann}}]{baguet_hartree-fock_2013}%
  \BibitemOpen
  \bibfield  {author} {\bibinfo {author} {\bibfnamefont {L.}~\bibnamefont
  {Baguet}}, \bibinfo {author} {\bibfnamefont {F.}~\bibnamefont {Delyon}},
  \bibinfo {author} {\bibfnamefont {B.}~\bibnamefont {Bernu}}, \ and\ \bibinfo
  {author} {\bibfnamefont {M.}~\bibnamefont {Holzmann}},\ }\href@noop {}
  {\bibfield  {journal} {\bibinfo  {journal} {Phys. Rev. Lett.}\ }\textbf
  {\bibinfo {volume} {111}},\ \bibinfo {pages} {166402} (\bibinfo {year}
  {2013})}\BibitemShut {NoStop}%
\bibitem [{\citenamefont {Zhang}\ and\ \citenamefont
  {Ceperley}(2008)}]{zhang_hartree-fock_2008}%
  \BibitemOpen
  \bibfield  {author} {\bibinfo {author} {\bibfnamefont {S.}~\bibnamefont
  {Zhang}}\ and\ \bibinfo {author} {\bibfnamefont {D.~M.}\ \bibnamefont
  {Ceperley}},\ }\href@noop {} {\bibfield  {journal} {\bibinfo  {journal}
  {Phys. Rev. Lett.}\ }\textbf {\bibinfo {volume} {100}},\ \bibinfo {pages}
  {236404} (\bibinfo {year} {2008})}\BibitemShut {NoStop}%
\bibitem [{\citenamefont {Gori-Giorgi}\ \emph {et~al.}(2000)\citenamefont
  {Gori-Giorgi}, \citenamefont {Sacchetti},\ and\ \citenamefont
  {Bachelet}}]{gori-giorgi_analytic_2000}%
  \BibitemOpen
  \bibfield  {author} {\bibinfo {author} {\bibfnamefont {P.}~\bibnamefont
  {Gori-Giorgi}}, \bibinfo {author} {\bibfnamefont {F.}~\bibnamefont
  {Sacchetti}}, \ and\ \bibinfo {author} {\bibfnamefont {G.~B.}\ \bibnamefont
  {Bachelet}},\ }\href@noop {} {\bibfield  {journal} {\bibinfo  {journal}
  {Phys. Rev. B}\ }\textbf {\bibinfo {volume} {61}},\ \bibinfo {pages} {7353}
  (\bibinfo {year} {2000})}\BibitemShut {NoStop}%
\bibitem [{\citenamefont {Gori-Giorgi}\ and\ \citenamefont
  {Perdew}(2001)}]{gori-giorgi_short-range_2001}%
  \BibitemOpen
  \bibfield  {author} {\bibinfo {author} {\bibfnamefont {P.}~\bibnamefont
  {Gori-Giorgi}}\ and\ \bibinfo {author} {\bibfnamefont {J.~P.}\ \bibnamefont
  {Perdew}},\ }\href@noop {} {\bibfield  {journal} {\bibinfo  {journal} {Phys.
  Rev. B}\ }\textbf {\bibinfo {volume} {64}},\ \bibinfo {pages} {155102}
  (\bibinfo {year} {2001})}\BibitemShut {NoStop}%
\bibitem [{\citenamefont {Petersilka}\ \emph {et~al.}(1996)\citenamefont
  {Petersilka}, \citenamefont {Gossmann},\ and\ \citenamefont
  {Gross}}]{petersilka_excitation_1996}%
  \BibitemOpen
  \bibfield  {author} {\bibinfo {author} {\bibfnamefont {M.}~\bibnamefont
  {Petersilka}}, \bibinfo {author} {\bibfnamefont {U.~J.}\ \bibnamefont
  {Gossmann}}, \ and\ \bibinfo {author} {\bibfnamefont {E.~K.~U.}\ \bibnamefont
  {Gross}},\ }\href@noop {} {\bibfield  {journal} {\bibinfo  {journal} {Phys.
  Rev. Lett.}\ }\textbf {\bibinfo {volume} {76}},\ \bibinfo {pages} {1212}
  (\bibinfo {year} {1996})}\BibitemShut {NoStop}%
\bibitem [{\citenamefont {Onida}\ \emph {et~al.}(2002)\citenamefont {Onida},
  \citenamefont {Reining},\ and\ \citenamefont
  {Rubio}}]{onida_electronic_2002}%
  \BibitemOpen
  \bibfield  {author} {\bibinfo {author} {\bibfnamefont {G.}~\bibnamefont
  {Onida}}, \bibinfo {author} {\bibfnamefont {L.}~\bibnamefont {Reining}}, \
  and\ \bibinfo {author} {\bibfnamefont {A.}~\bibnamefont {Rubio}},\
  }\href@noop {} {\bibfield  {journal} {\bibinfo  {journal} {Rev. Mod. Phys.}\
  }\textbf {\bibinfo {volume} {74}},\ \bibinfo {pages} {601} (\bibinfo {year}
  {2002})}\BibitemShut {NoStop}%
\bibitem [{\citenamefont {Hedin}(1965)}]{hedin_new_1965}%
  \BibitemOpen
  \bibfield  {author} {\bibinfo {author} {\bibfnamefont {L.}~\bibnamefont
  {Hedin}},\ }\href@noop {} {\bibfield  {journal} {\bibinfo  {journal} {Phys.
  Rev.}\ }\textbf {\bibinfo {volume} {139}},\ \bibinfo {pages} {A796} (\bibinfo
  {year} {1965})}\BibitemShut {NoStop}%
\bibitem [{\citenamefont {Almbladh}\ \emph {et~al.}(1999)\citenamefont
  {Almbladh}, \citenamefont {von Barth},\ and\ \citenamefont {van
  Leeuwen}}]{almbladh_variational_1999}%
  \BibitemOpen
  \bibfield  {author} {\bibinfo {author} {\bibfnamefont {C.-O.}\ \bibnamefont
  {Almbladh}}, \bibinfo {author} {\bibfnamefont {U.}~\bibnamefont {von Barth}},
  \ and\ \bibinfo {author} {\bibfnamefont {R.}~\bibnamefont {van Leeuwen}},\
  }\href@noop {} {\bibfield  {journal} {\bibinfo  {journal} {International
  Journal of Modern Physics B}\ }\textbf {\bibinfo {volume} {13}},\ \bibinfo
  {pages} {535} (\bibinfo {year} {1999})}\BibitemShut {NoStop}%
\bibitem [{\citenamefont {Garc{\'\i}a-Gonz{\'a}lez}\ and\ \citenamefont
  {Godby}(2001)}]{garcia-gonzalez_self-consistent_2001}%
  \BibitemOpen
  \bibfield  {author} {\bibinfo {author} {\bibfnamefont {P.}~\bibnamefont
  {Garc{\'\i}a-Gonz{\'a}lez}}\ and\ \bibinfo {author} {\bibfnamefont {R.~W.}\
  \bibnamefont {Godby}},\ }\href@noop {} {\bibfield  {journal} {\bibinfo
  {journal} {Physical Review B}\ }\textbf {\bibinfo {volume} {63}},\ \bibinfo
  {pages} {075112} (\bibinfo {year} {2001})}\BibitemShut {NoStop}%
\bibitem [{\citenamefont {Dahlen}\ and\ \citenamefont {van
  Leeuwen}(2005)}]{dahlen_self-consistent_2005}%
  \BibitemOpen
  \bibfield  {author} {\bibinfo {author} {\bibfnamefont {N.~E.}\ \bibnamefont
  {Dahlen}}\ and\ \bibinfo {author} {\bibfnamefont {R.}~\bibnamefont {van
  Leeuwen}},\ }\href@noop {} {\bibfield  {journal} {\bibinfo  {journal} {J.
  Chem. Phys.}\ }\textbf {\bibinfo {volume} {122}},\ \bibinfo {pages} {164102}
  (\bibinfo {year} {2005})}\BibitemShut {NoStop}%
\bibitem [{\citenamefont {Stan}\ \emph {et~al.}(2009)\citenamefont {Stan},
  \citenamefont {Dahlen},\ and\ \citenamefont {van
  Leeuwen}}]{stan_levels_2009}%
  \BibitemOpen
  \bibfield  {author} {\bibinfo {author} {\bibfnamefont {A.}~\bibnamefont
  {Stan}}, \bibinfo {author} {\bibfnamefont {N.~E.}\ \bibnamefont {Dahlen}}, \
  and\ \bibinfo {author} {\bibfnamefont {R.}~\bibnamefont {van Leeuwen}},\
  }\href@noop {} {\bibfield  {journal} {\bibinfo  {journal} {J. Chem. Phys.}\
  }\textbf {\bibinfo {volume} {130}},\ \bibinfo {pages} {114105} (\bibinfo
  {year} {2009})}\BibitemShut {NoStop}%
\bibitem [{\citenamefont
  {Lundqvist}(1967{\natexlab{a}})}]{lundqvist_single-particle_1967}%
  \BibitemOpen
  \bibfield  {author} {\bibinfo {author} {\bibfnamefont {B.~I.}\ \bibnamefont
  {Lundqvist}},\ }\href@noop {} {\bibfield  {journal} {\bibinfo  {journal}
  {Phys. Kondens. Mater.}\ }\textbf {\bibinfo {volume} {6}},\ \bibinfo {pages}
  {193} (\bibinfo {year} {1967}{\natexlab{a}})}\BibitemShut {NoStop}%
\bibitem [{\citenamefont
  {Lundqvist}(1967{\natexlab{b}})}]{lundqvist_single-particle_1967-1}%
  \BibitemOpen
  \bibfield  {author} {\bibinfo {author} {\bibfnamefont {B.~I.}\ \bibnamefont
  {Lundqvist}},\ }\href@noop {} {\bibfield  {journal} {\bibinfo  {journal}
  {Phys. Kondens. Mater.}\ }\textbf {\bibinfo {volume} {6}},\ \bibinfo {pages}
  {206} (\bibinfo {year} {1967}{\natexlab{b}})}\BibitemShut {NoStop}%
\bibitem [{\citenamefont {Lundqvist}(1968)}]{lundqvist_single-particle_1968}%
  \BibitemOpen
  \bibfield  {author} {\bibinfo {author} {\bibfnamefont {B.~I.}\ \bibnamefont
  {Lundqvist}},\ }\href@noop {} {\bibfield  {journal} {\bibinfo  {journal}
  {Phys. Kondens. Mater.}\ }\textbf {\bibinfo {volume} {7}},\ \bibinfo {pages}
  {117} (\bibinfo {year} {1968})}\BibitemShut {NoStop}%
\bibitem [{\citenamefont {Holm}\ and\ \citenamefont {von
  Barth}(1998)}]{holm_fully_1998}%
  \BibitemOpen
  \bibfield  {author} {\bibinfo {author} {\bibfnamefont {B.}~\bibnamefont
  {Holm}}\ and\ \bibinfo {author} {\bibfnamefont {U.}~\bibnamefont {von
  Barth}},\ }\href@noop {} {\bibfield  {journal} {\bibinfo  {journal} {Phys.
  Rev. B}\ }\textbf {\bibinfo {volume} {57}},\ \bibinfo {pages} {2108}
  (\bibinfo {year} {1998})}\BibitemShut {NoStop}%
\bibitem [{\citenamefont {Yasuhara}\ \emph {et~al.}(1999)\citenamefont
  {Yasuhara}, \citenamefont {Yoshinaga},\ and\ \citenamefont
  {Higuchi}}]{yasuhara_why_1999}%
  \BibitemOpen
  \bibfield  {author} {\bibinfo {author} {\bibfnamefont {H.}~\bibnamefont
  {Yasuhara}}, \bibinfo {author} {\bibfnamefont {S.}~\bibnamefont {Yoshinaga}},
  \ and\ \bibinfo {author} {\bibfnamefont {M.}~\bibnamefont {Higuchi}},\
  }\href@noop {} {\bibfield  {journal} {\bibinfo  {journal} {Phys. Rev. Lett.}\
  }\textbf {\bibinfo {volume} {83}},\ \bibinfo {pages} {3250} (\bibinfo {year}
  {1999})}\BibitemShut {NoStop}%
\bibitem [{\citenamefont {Takada}(2001)}]{takada_inclusion_2001}%
  \BibitemOpen
  \bibfield  {author} {\bibinfo {author} {\bibfnamefont {Y.}~\bibnamefont
  {Takada}},\ }\href@noop {} {\bibfield  {journal} {\bibinfo  {journal} {Phys.
  Rev. Lett.}\ }\textbf {\bibinfo {volume} {87}},\ \bibinfo {pages} {226402}
  (\bibinfo {year} {2001})}\BibitemShut {NoStop}%
\bibitem [{\citenamefont {Kwong}\ and\ \citenamefont
  {Bonitz}(2000)}]{kwong_real-time_2000}%
  \BibitemOpen
  \bibfield  {author} {\bibinfo {author} {\bibfnamefont {N.-H.}\ \bibnamefont
  {Kwong}}\ and\ \bibinfo {author} {\bibfnamefont {M.}~\bibnamefont {Bonitz}},\
  }\href@noop {} {\bibfield  {journal} {\bibinfo  {journal} {Phys. Rev. Lett.}\
  }\textbf {\bibinfo {volume} {84}},\ \bibinfo {pages} {1768} (\bibinfo {year}
  {2000})}\BibitemShut {NoStop}%
\bibitem [{\citenamefont {Pal}\ \emph {et~al.}(2009)\citenamefont {Pal},
  \citenamefont {Pavlyukh}, \citenamefont {Schneider},\ and\ \citenamefont
  {H\"{u}bner}}]{pal_conserving_2009}%
  \BibitemOpen
  \bibfield  {author} {\bibinfo {author} {\bibfnamefont {G.}~\bibnamefont
  {Pal}}, \bibinfo {author} {\bibfnamefont {Y.}~\bibnamefont {Pavlyukh}},
  \bibinfo {author} {\bibfnamefont {H.~C.}\ \bibnamefont {Schneider}}, \ and\
  \bibinfo {author} {\bibfnamefont {W.}~\bibnamefont {H\"{u}bner}},\
  }\href@noop {} {\bibfield  {journal} {\bibinfo  {journal} {Eur. Phys. J. B}\
  }\textbf {\bibinfo {volume} {70}},\ \bibinfo {pages} {483} (\bibinfo {year}
  {2009})}\BibitemShut {NoStop}%
\bibitem [{\citenamefont {Mahan}\ and\ \citenamefont
  {Sernelius}(1989)}]{mahan_electron-electron_1989}%
  \BibitemOpen
  \bibfield  {author} {\bibinfo {author} {\bibfnamefont {G.~D.}\ \bibnamefont
  {Mahan}}\ and\ \bibinfo {author} {\bibfnamefont {B.~E.}\ \bibnamefont
  {Sernelius}},\ }\href@noop {} {\bibfield  {journal} {\bibinfo  {journal}
  {Phys. Rev. Lett.}\ }\textbf {\bibinfo {volume} {62}},\ \bibinfo {pages}
  {2718} (\bibinfo {year} {1989})}\BibitemShut {NoStop}%
\bibitem [{\citenamefont {Mahan}(1994)}]{mahan_gw_1994}%
  \BibitemOpen
  \bibfield  {author} {\bibinfo {author} {\bibfnamefont {G.}~\bibnamefont
  {Mahan}},\ }in\ \href@noop {} {\emph {\bibinfo {booktitle} {Comments on
  {Condensed} {Matter} {Physics}}}},\ \bibinfo {series} {Comments on {Modern}
  {Physics}}, Vol.~\bibinfo {volume} {16}\ (\bibinfo  {publisher} {Gordon and
  Breach Science Publishers, SA},\ \bibinfo {year} {1994})\ p.\ \bibinfo
  {pages} {333}\BibitemShut {NoStop}%
\bibitem [{\citenamefont {Bobbert}\ and\ \citenamefont {van
  Haeringen}(1994)}]{bobbert_lowest-order_1994}%
  \BibitemOpen
  \bibfield  {author} {\bibinfo {author} {\bibfnamefont {P.~A.}\ \bibnamefont
  {Bobbert}}\ and\ \bibinfo {author} {\bibfnamefont {W.}~\bibnamefont {van
  Haeringen}},\ }\href@noop {} {\bibfield  {journal} {\bibinfo  {journal}
  {Phys. Rev. B}\ }\textbf {\bibinfo {volume} {49}},\ \bibinfo {pages} {10326}
  (\bibinfo {year} {1994})}\BibitemShut {NoStop}%
\bibitem [{\citenamefont {Schindlmayr}\ and\ \citenamefont
  {Godby}(1998)}]{schindlmayr_systematic_1998}%
  \BibitemOpen
  \bibfield  {author} {\bibinfo {author} {\bibfnamefont {A.}~\bibnamefont
  {Schindlmayr}}\ and\ \bibinfo {author} {\bibfnamefont {R.~W.}\ \bibnamefont
  {Godby}},\ }\href@noop {} {\bibfield  {journal} {\bibinfo  {journal} {Phys.
  Rev. Lett.}\ }\textbf {\bibinfo {volume} {80}},\ \bibinfo {pages} {1702}
  (\bibinfo {year} {1998})}\BibitemShut {NoStop}%
\bibitem [{\citenamefont {Minnhagen}(1975)}]{minnhagen_aspects_1975}%
  \BibitemOpen
  \bibfield  {author} {\bibinfo {author} {\bibfnamefont {P.}~\bibnamefont
  {Minnhagen}},\ }\href@noop {} {\bibfield  {journal} {\bibinfo  {journal} {J.
  Phys. C}\ }\textbf {\bibinfo {volume} {8}},\ \bibinfo {pages} {1535}
  (\bibinfo {year} {1975})}\BibitemShut {NoStop}%
\bibitem [{\citenamefont {Shirley}(1996)}]{shirley_self-consistent_1996}%
  \BibitemOpen
  \bibfield  {author} {\bibinfo {author} {\bibfnamefont {E.~L.}\ \bibnamefont
  {Shirley}},\ }\href@noop {} {\bibfield  {journal} {\bibinfo  {journal} {Phys.
  Rev. B}\ }\textbf {\bibinfo {volume} {54}},\ \bibinfo {pages} {7758}
  (\bibinfo {year} {1996})}\BibitemShut {NoStop}%
\bibitem [{\citenamefont {Takada}\ and\ \citenamefont
  {Yasuhara}(2002)}]{takada_dynamical_2002}%
  \BibitemOpen
  \bibfield  {author} {\bibinfo {author} {\bibfnamefont {Y.}~\bibnamefont
  {Takada}}\ and\ \bibinfo {author} {\bibfnamefont {H.}~\bibnamefont
  {Yasuhara}},\ }\href@noop {} {\bibfield  {journal} {\bibinfo  {journal}
  {Phys. Rev. Lett.}\ }\textbf {\bibinfo {volume} {89}},\ \bibinfo {pages}
  {216402} (\bibinfo {year} {2002})}\BibitemShut {NoStop}%
\bibitem [{\citenamefont {Bruneval}\ \emph {et~al.}(2005)\citenamefont
  {Bruneval}, \citenamefont {Sottile}, \citenamefont {Olevano}, \citenamefont
  {Del~Sole},\ and\ \citenamefont {Reining}}]{bruneval_many-body_2005}%
  \BibitemOpen
  \bibfield  {author} {\bibinfo {author} {\bibfnamefont {F.}~\bibnamefont
  {Bruneval}}, \bibinfo {author} {\bibfnamefont {F.}~\bibnamefont {Sottile}},
  \bibinfo {author} {\bibfnamefont {V.}~\bibnamefont {Olevano}}, \bibinfo
  {author} {\bibfnamefont {R.}~\bibnamefont {Del~Sole}}, \ and\ \bibinfo
  {author} {\bibfnamefont {L.}~\bibnamefont {Reining}},\ }\href@noop {}
  {\bibfield  {journal} {\bibinfo  {journal} {Phys. Rev. Lett.}\ }\textbf
  {\bibinfo {volume} {94}},\ \bibinfo {pages} {186402} (\bibinfo {year}
  {2005})}\BibitemShut {NoStop}%
\bibitem [{\citenamefont {Maebashi}\ and\ \citenamefont
  {Takada}(2011)}]{maebashi_analysis_2011}%
  \BibitemOpen
  \bibfield  {author} {\bibinfo {author} {\bibfnamefont {H.}~\bibnamefont
  {Maebashi}}\ and\ \bibinfo {author} {\bibfnamefont {Y.}~\bibnamefont
  {Takada}},\ }\href@noop {} {\bibfield  {journal} {\bibinfo  {journal} {Phys.
  Rev. B}\ }\textbf {\bibinfo {volume} {84}},\ \bibinfo {pages} {245134}
  (\bibinfo {year} {2011})}\BibitemShut {NoStop}%
\bibitem [{\citenamefont {Holm}\ and\ \citenamefont
  {Aryasetiawan}(1997)}]{holm_self-consistent_1997}%
  \BibitemOpen
  \bibfield  {author} {\bibinfo {author} {\bibfnamefont {B.}~\bibnamefont
  {Holm}}\ and\ \bibinfo {author} {\bibfnamefont {F.}~\bibnamefont
  {Aryasetiawan}},\ }\href@noop {} {\bibfield  {journal} {\bibinfo  {journal}
  {Phys. Rev. B}\ }\textbf {\bibinfo {volume} {56}},\ \bibinfo {pages} {12825}
  (\bibinfo {year} {1997})}\BibitemShut {NoStop}%
\bibitem [{\citenamefont {Kas}\ \emph {et~al.}(2014)\citenamefont {Kas},
  \citenamefont {Rehr},\ and\ \citenamefont {Reining}}]{kas_cumulant_2014}%
  \BibitemOpen
  \bibfield  {author} {\bibinfo {author} {\bibfnamefont {J.~J.}\ \bibnamefont
  {Kas}}, \bibinfo {author} {\bibfnamefont {J.~J.}\ \bibnamefont {Rehr}}, \
  and\ \bibinfo {author} {\bibfnamefont {L.}~\bibnamefont {Reining}},\
  }\href@noop {} {\bibfield  {journal} {\bibinfo  {journal} {Phys. Rev. B}\
  }\textbf {\bibinfo {volume} {90}},\ \bibinfo {pages} {085112} (\bibinfo
  {year} {2014})}\BibitemShut {NoStop}%
\bibitem [{\citenamefont {Caruso}\ and\ \citenamefont
  {Giustino}(2016)}]{caruso_cumulant_2016}%
  \BibitemOpen
  \bibfield  {author} {\bibinfo {author} {\bibfnamefont {F.}~\bibnamefont
  {Caruso}}\ and\ \bibinfo {author} {\bibfnamefont {F.}~\bibnamefont
  {Giustino}},\ }\href@noop {} {\bibfield  {journal} {\bibinfo  {journal}
  {arXiv:1606.08573}\ } (\bibinfo {year} {2016})}\BibitemShut {NoStop}%
\bibitem [{Note1()}]{Note1}%
  \BibitemOpen
  \bibinfo {note} {For instance the cumulant expansion is exact for deep core
  states interacting with plasmons and leads to the spectrum with equally
  spaced satellites~\cite {langreth_singularities_1970}. Yet, this assumption
  is less justified for the valence band excitations overestimating the weight
  of higher order plasmon satellites (something that can be partially cured by
  taking multiple plasmon branches and their dispersion into account~\cite
  {cini_coherent_1986,guzzo_multiple_2014}).}\BibitemShut {Stop}%
\bibitem [{\citenamefont {Minnhagen}(1974)}]{minnhagen_vertex_1974}%
  \BibitemOpen
  \bibfield  {author} {\bibinfo {author} {\bibfnamefont {P.}~\bibnamefont
  {Minnhagen}},\ }\href@noop {} {\bibfield  {journal} {\bibinfo  {journal} {J.
  Phys. C}\ }\textbf {\bibinfo {volume} {7}},\ \bibinfo {pages} {3013}
  (\bibinfo {year} {1974})}\BibitemShut {NoStop}%
\bibitem [{PSD()}]{PSDtot}%
  \BibitemOpen
  \href@noop {} {}\bibinfo {note} {G. Stefanucci, Y. Pavlyukh, A.-M. Uimonen,
  and R. van Leeuwen, Phys. Rev. B {\bf 90}, 115134 (2014); {\em ibid} {\bf
  93}, 119906(E) (2016).}\BibitemShut {Stop}%
\bibitem [{\citenamefont {Uimonen}\ \emph {et~al.}(2015)\citenamefont
  {Uimonen}, \citenamefont {Stefanucci}, \citenamefont {Pavlyukh},\ and\
  \citenamefont {van Leeuwen}}]{uimonen_diagrammatic_2015}%
  \BibitemOpen
  \bibfield  {author} {\bibinfo {author} {\bibfnamefont {A.-M.}\ \bibnamefont
  {Uimonen}}, \bibinfo {author} {\bibfnamefont {G.}~\bibnamefont {Stefanucci}},
  \bibinfo {author} {\bibfnamefont {Y.}~\bibnamefont {Pavlyukh}}, \ and\
  \bibinfo {author} {\bibfnamefont {R.}~\bibnamefont {van Leeuwen}},\
  }\href@noop {} {\bibfield  {journal} {\bibinfo  {journal} {Phys. Rev. B}\
  }\textbf {\bibinfo {volume} {91}},\ \bibinfo {pages} {115104} (\bibinfo
  {year} {2015})}\BibitemShut {NoStop}%
\bibitem [{\citenamefont {Karlsson}\ and\ \citenamefont {van
  Leeuwen}(2016)}]{KvL.2016}%
  \BibitemOpen
  \bibfield  {author} {\bibinfo {author} {\bibfnamefont {D.}~\bibnamefont
  {Karlsson}}\ and\ \bibinfo {author} {\bibfnamefont {R.}~\bibnamefont {van
  Leeuwen}},\ }\href@noop {} {\bibfield  {journal} {\bibinfo  {journal}
  {arxiv:1606.07486 [cond-mat]}\ } (\bibinfo {year} {2016})}\BibitemShut
  {NoStop}%
\bibitem [{Note2()}]{Note2}%
  \BibitemOpen
  \bibinfo {note} {We start with zeroth approximation $\Delta \mu
  ^{(0)}=\protect \text {Re}\protect \tmspace +\thinmuskip {.1667em}\Sigma
  (k_F,1/2k_F^2)$ and perform two more calculations for $k=k_F\pm \Delta k$,
  where $\Delta k$ is a small number, typically a few percents of the Fermi
  momentum. The refined chemical potential shift is then given by $\Delta \mu
  =\Delta \mu ^{(0)}+\protect \frac 12(\Delta \epsilon _{k_F+\Delta k}+\Delta
  \epsilon _{k_F-\Delta k})$, where $\Delta \epsilon _{k}$ is the correlational
  shift.}\BibitemShut {Stop}%
\bibitem [{\citenamefont {Lyo}\ and\ \citenamefont
  {Plummer}(1988)}]{lyo_quasiparticle_1988}%
  \BibitemOpen
  \bibfield  {author} {\bibinfo {author} {\bibfnamefont {I.-W.}\ \bibnamefont
  {Lyo}}\ and\ \bibinfo {author} {\bibfnamefont {E.~W.}\ \bibnamefont
  {Plummer}},\ }\href@noop {} {\bibfield  {journal} {\bibinfo  {journal} {Phys.
  Rev. Lett.}\ }\textbf {\bibinfo {volume} {60}},\ \bibinfo {pages} {1558}
  (\bibinfo {year} {1988})}\BibitemShut {NoStop}%
\bibitem [{\citenamefont {Aryasetiawan}\ \emph {et~al.}(1996)\citenamefont
  {Aryasetiawan}, \citenamefont {Hedin},\ and\ \citenamefont
  {Karlsson}}]{aryasetiawan_multiple_1996}%
  \BibitemOpen
  \bibfield  {author} {\bibinfo {author} {\bibfnamefont {F.}~\bibnamefont
  {Aryasetiawan}}, \bibinfo {author} {\bibfnamefont {L.}~\bibnamefont {Hedin}},
  \ and\ \bibinfo {author} {\bibfnamefont {K.}~\bibnamefont {Karlsson}},\
  }\href@noop {} {\bibfield  {journal} {\bibinfo  {journal} {Phys. Rev. Lett.}\
  }\textbf {\bibinfo {volume} {77}},\ \bibinfo {pages} {2268} (\bibinfo {year}
  {1996})}\BibitemShut {NoStop}%
\bibitem [{Note3()}]{Note3}%
  \BibitemOpen
  \bibinfo {note} {If $W$ in $\Sigma _{a\protect \mathaccentV {bar}184{a}}$ is
  replaced by the bare Coulomb interaction the so-called second-order exchange
  SE is obtained. Its on-shell value is density independent and it is known
  analytically~\cite
  {onsager_integrals_1966,ziesche_self-energy_2007,glasser_analysis_2007},
  $\Sigma _{2x}(k_F,1/2k_F^2)={\setbox \z@ \hbox {\frozen@everymath \@emptytoks
  \mathsurround \z@ $\nulldelimiterspace \z@ \left [\vcenter to1.2\big@size
  {}\right .$}\box \z@ }2\pi ^2\protect \qopname \relax o{ln}(2)/3-3\zeta
  (3){\setbox \z@ \hbox {\frozen@everymath \@emptytoks \mathsurround \z@
  $\nulldelimiterspace \z@ \left ]\vcenter to1.2\big@size {}\right .$}\box \z@
  }/4\pi ^2$. This result represents a useful check for our numerical
  algorithms.}\BibitemShut {Stop}%
\bibitem [{\citenamefont {Pavlyukh}\ \emph {et~al.}(2013)\citenamefont
  {Pavlyukh}, \citenamefont {Berakdar},\ and\ \citenamefont
  {Rubio}}]{pavlyukh_initial_2013}%
  \BibitemOpen
  \bibfield  {author} {\bibinfo {author} {\bibfnamefont {Y.}~\bibnamefont
  {Pavlyukh}}, \bibinfo {author} {\bibfnamefont {J.}~\bibnamefont {Berakdar}},
  \ and\ \bibinfo {author} {\bibfnamefont {A.}~\bibnamefont {Rubio}},\
  }\href@noop {} {\bibfield  {journal} {\bibinfo  {journal} {Phys. Rev. B}\
  }\textbf {\bibinfo {volume} {87}},\ \bibinfo {pages} {125101} (\bibinfo
  {year} {2013})}\BibitemShut {NoStop}%
\bibitem [{\citenamefont {Mahaux}\ and\ \citenamefont
  {Sartor}(1992)}]{mahaux_theoretical_1992}%
  \BibitemOpen
  \bibfield  {author} {\bibinfo {author} {\bibfnamefont {C.}~\bibnamefont
  {Mahaux}}\ and\ \bibinfo {author} {\bibfnamefont {R.}~\bibnamefont
  {Sartor}},\ }\href@noop {} {\bibfield  {journal} {\bibinfo  {journal} {Phys.
  Rep.}\ }\textbf {\bibinfo {volume} {211}},\ \bibinfo {pages} {53} (\bibinfo
  {year} {1992})}\BibitemShut {NoStop}%
\bibitem [{\citenamefont {Gori-Giorgi}\ and\ \citenamefont
  {Ziesche}(2002)}]{gori_giorgi_momentum_2002}%
  \BibitemOpen
  \bibfield  {author} {\bibinfo {author} {\bibfnamefont {P.}~\bibnamefont
  {Gori-Giorgi}}\ and\ \bibinfo {author} {\bibfnamefont {P.}~\bibnamefont
  {Ziesche}},\ }\href@noop {} {\bibfield  {journal} {\bibinfo  {journal} {Phys.
  Rev. B}\ }\textbf {\bibinfo {volume} {66}},\ \bibinfo {pages} {235116}
  (\bibinfo {year} {2002})}\BibitemShut {NoStop}%
\bibitem [{\citenamefont {Olevano}\ \emph {et~al.}(2012)\citenamefont
  {Olevano}, \citenamefont {Titov}, \citenamefont {Ladisa}, \citenamefont
  {H{\"a}m{\"a}l{\"a}inen}, \citenamefont {Huotari},\ and\ \citenamefont
  {Holzmann}}]{olevano_momentum_2012}%
  \BibitemOpen
  \bibfield  {author} {\bibinfo {author} {\bibfnamefont {V.}~\bibnamefont
  {Olevano}}, \bibinfo {author} {\bibfnamefont {A.}~\bibnamefont {Titov}},
  \bibinfo {author} {\bibfnamefont {M.}~\bibnamefont {Ladisa}}, \bibinfo
  {author} {\bibfnamefont {K.}~\bibnamefont {H{\"a}m{\"a}l{\"a}inen}}, \bibinfo
  {author} {\bibfnamefont {S.}~\bibnamefont {Huotari}}, \ and\ \bibinfo
  {author} {\bibfnamefont {M.}~\bibnamefont {Holzmann}},\ }\href@noop {}
  {\bibfield  {journal} {\bibinfo  {journal} {Phys. Rev. B}\ }\textbf {\bibinfo
  {volume} {86}},\ \bibinfo {pages} {195123} (\bibinfo {year}
  {2012})}\BibitemShut {NoStop}%
\bibitem [{\citenamefont {Echenique}\ \emph {et~al.}(2000)\citenamefont
  {Echenique}, \citenamefont {Pitarke}, \citenamefont {Chulkov},\ and\
  \citenamefont {Rubio}}]{echenique_theory_2000}%
  \BibitemOpen
  \bibfield  {author} {\bibinfo {author} {\bibfnamefont {P.~M.}\ \bibnamefont
  {Echenique}}, \bibinfo {author} {\bibfnamefont {J.~M.}\ \bibnamefont
  {Pitarke}}, \bibinfo {author} {\bibfnamefont {E.~V.}\ \bibnamefont
  {Chulkov}}, \ and\ \bibinfo {author} {\bibfnamefont {A.}~\bibnamefont
  {Rubio}},\ }\href@noop {} {\bibfield  {journal} {\bibinfo  {journal} {Chem.
  Phys.}\ }\textbf {\bibinfo {volume} {251}},\ \bibinfo {pages} {1} (\bibinfo
  {year} {2000})}\BibitemShut {NoStop}%
\bibitem [{\citenamefont {Qian}\ and\ \citenamefont
  {Vignale}(2005)}]{qian_lifetime_2005}%
  \BibitemOpen
  \bibfield  {author} {\bibinfo {author} {\bibfnamefont {Z.}~\bibnamefont
  {Qian}}\ and\ \bibinfo {author} {\bibfnamefont {G.}~\bibnamefont {Vignale}},\
  }\href@noop {} {\bibfield  {journal} {\bibinfo  {journal} {Phys. Rev. B}\
  }\textbf {\bibinfo {volume} {71}},\ \bibinfo {pages} {075112} (\bibinfo
  {year} {2005})}\BibitemShut {NoStop}%
\bibitem [{\citenamefont {Zhukov}\ \emph {et~al.}(2002)\citenamefont {Zhukov},
  \citenamefont {Aryasetiawan}, \citenamefont {Chulkov},\ and\ \citenamefont
  {Echenique}}]{zhukov_lifetimes_2002}%
  \BibitemOpen
  \bibfield  {author} {\bibinfo {author} {\bibfnamefont {V.~P.}\ \bibnamefont
  {Zhukov}}, \bibinfo {author} {\bibfnamefont {F.}~\bibnamefont
  {Aryasetiawan}}, \bibinfo {author} {\bibfnamefont {E.~V.}\ \bibnamefont
  {Chulkov}}, \ and\ \bibinfo {author} {\bibfnamefont {P.~M.}\ \bibnamefont
  {Echenique}},\ }\href@noop {} {\bibfield  {journal} {\bibinfo  {journal}
  {Phys. Rev. B}\ }\textbf {\bibinfo {volume} {65}},\ \bibinfo {pages} {115116}
  (\bibinfo {year} {2002})}\BibitemShut {NoStop}%
\bibitem [{\citenamefont {Pavlyukh}\ \emph {et~al.}(2008)\citenamefont
  {Pavlyukh}, \citenamefont {Berakdar},\ and\ \citenamefont
  {H\"{u}bner}}]{pavlyukh_decay_2008}%
  \BibitemOpen
  \bibfield  {author} {\bibinfo {author} {\bibfnamefont {Y.}~\bibnamefont
  {Pavlyukh}}, \bibinfo {author} {\bibfnamefont {J.}~\bibnamefont {Berakdar}},
  \ and\ \bibinfo {author} {\bibfnamefont {W.}~\bibnamefont {H\"{u}bner}},\
  }\href@noop {} {\bibfield  {journal} {\bibinfo  {journal} {Phys. Rev. Lett.}\
  }\textbf {\bibinfo {volume} {100}},\ \bibinfo {pages} {116103} (\bibinfo
  {year} {2008})}\BibitemShut {NoStop}%
\bibitem [{\citenamefont {Pavlyukh}\ and\ \citenamefont
  {Berakdar}(2011)}]{pavlyukh_communication:_2011}%
  \BibitemOpen
  \bibfield  {author} {\bibinfo {author} {\bibfnamefont {Y.}~\bibnamefont
  {Pavlyukh}}\ and\ \bibinfo {author} {\bibfnamefont {J.}~\bibnamefont
  {Berakdar}},\ }\href@noop {} {\bibfield  {journal} {\bibinfo  {journal} {J.
  Chem. Phys.}\ }\textbf {\bibinfo {volume} {135}},\ \bibinfo {pages} {201103}
  (\bibinfo {year} {2011})}\BibitemShut {NoStop}%
\bibitem [{\citenamefont {Nechaev}\ and\ \citenamefont
  {Chulkov}(2005)}]{nechaev_variational_2005}%
  \BibitemOpen
  \bibfield  {author} {\bibinfo {author} {\bibfnamefont {I.~A.}\ \bibnamefont
  {Nechaev}}\ and\ \bibinfo {author} {\bibfnamefont {E.~V.}\ \bibnamefont
  {Chulkov}},\ }\href@noop {} {\bibfield  {journal} {\bibinfo  {journal} {Phys.
  Rev. B}\ }\textbf {\bibinfo {volume} {71}},\ \bibinfo {pages} {115104}
  (\bibinfo {year} {2005})}\BibitemShut {NoStop}%
\bibitem [{\citenamefont {Verdozzi}\ \emph {et~al.}(1995)\citenamefont
  {Verdozzi}, \citenamefont {Godby},\ and\ \citenamefont
  {Holloway}}]{verdozzi_evaluation_1995}%
  \BibitemOpen
  \bibfield  {author} {\bibinfo {author} {\bibfnamefont {C.}~\bibnamefont
  {Verdozzi}}, \bibinfo {author} {\bibfnamefont {R.~W.}\ \bibnamefont {Godby}},
  \ and\ \bibinfo {author} {\bibfnamefont {S.}~\bibnamefont {Holloway}},\
  }\href@noop {} {\bibfield  {journal} {\bibinfo  {journal} {Phys. Rev. Lett.}\
  }\textbf {\bibinfo {volume} {74}},\ \bibinfo {pages} {2327} (\bibinfo {year}
  {1995})}\BibitemShut {NoStop}%
\bibitem [{Note4()}]{Note4}%
  \BibitemOpen
  \bibinfo {note} {Due to the long-range character of the Coulomb interaction
  (bare and to a lesser extent screened one) the integrations need to be
  extended beyond the boundaries of the first Brillouin zone. For $GW$
  calculations the fast Fourier method have been proposed~\cite
  {rojas_space-time_1995}. For higher dimensional integrals, such as in the
  present calculations, this approach becomes impractical, but can be remedied
  by the Monte Carlo integration featuring excellent scalability.}\BibitemShut
  {Stop}%
\bibitem [{\citenamefont {Northrup}\ \emph {et~al.}(1987)\citenamefont
  {Northrup}, \citenamefont {Hybertsen},\ and\ \citenamefont
  {Louie}}]{northrup_theory_1987}%
  \BibitemOpen
  \bibfield  {author} {\bibinfo {author} {\bibfnamefont {J.~E.}\ \bibnamefont
  {Northrup}}, \bibinfo {author} {\bibfnamefont {M.~S.}\ \bibnamefont
  {Hybertsen}}, \ and\ \bibinfo {author} {\bibfnamefont {S.~G.}\ \bibnamefont
  {Louie}},\ }\href@noop {} {\bibfield  {journal} {\bibinfo  {journal} {Phys.
  Rev. Lett.}\ }\textbf {\bibinfo {volume} {59}},\ \bibinfo {pages} {819}
  (\bibinfo {year} {1987})}\BibitemShut {NoStop}%
\bibitem [{\citenamefont {Lischner}\ \emph {et~al.}(2014)\citenamefont
  {Lischner}, \citenamefont {Bazhirov}, \citenamefont {MacDonald},
  \citenamefont {Cohen},\ and\ \citenamefont {Louie}}]{lischner_effect_2014}%
  \BibitemOpen
  \bibfield  {author} {\bibinfo {author} {\bibfnamefont {J.}~\bibnamefont
  {Lischner}}, \bibinfo {author} {\bibfnamefont {T.}~\bibnamefont {Bazhirov}},
  \bibinfo {author} {\bibfnamefont {A.~H.}\ \bibnamefont {MacDonald}}, \bibinfo
  {author} {\bibfnamefont {M.~L.}\ \bibnamefont {Cohen}}, \ and\ \bibinfo
  {author} {\bibfnamefont {S.~G.}\ \bibnamefont {Louie}},\ }\href@noop {}
  {\bibfield  {journal} {\bibinfo  {journal} {Phys. Rev. B}\ }\textbf {\bibinfo
  {volume} {89}} (\bibinfo {year} {2014})}\BibitemShut {NoStop}%
\bibitem [{\citenamefont {Langreth}(1970)}]{langreth_singularities_1970}%
  \BibitemOpen
  \bibfield  {author} {\bibinfo {author} {\bibfnamefont {D.~C.}\ \bibnamefont
  {Langreth}},\ }\href@noop {} {\bibfield  {journal} {\bibinfo  {journal}
  {Phys. Rev. B}\ }\textbf {\bibinfo {volume} {1}},\ \bibinfo {pages} {471}
  (\bibinfo {year} {1970})}\BibitemShut {NoStop}%
\bibitem [{\citenamefont {Cini}(1986)}]{cini_coherent_1986}%
  \BibitemOpen
  \bibfield  {author} {\bibinfo {author} {\bibfnamefont {M.}~\bibnamefont
  {Cini}},\ }\href@noop {} {\bibfield  {journal} {\bibinfo  {journal} {Journal
  of Physics C: Solid State Physics}\ }\textbf {\bibinfo {volume} {19}},\
  \bibinfo {pages} {429} (\bibinfo {year} {1986})}\BibitemShut {NoStop}%
\bibitem [{\citenamefont {Guzzo}\ \emph {et~al.}(2014)\citenamefont {Guzzo},
  \citenamefont {Kas}, \citenamefont {Sponza}, \citenamefont {Giorgetti},
  \citenamefont {Sottile}, \citenamefont {Pierucci}, \citenamefont {Silly},
  \citenamefont {Sirotti}, \citenamefont {Rehr},\ and\ \citenamefont
  {Reining}}]{guzzo_multiple_2014}%
  \BibitemOpen
  \bibfield  {author} {\bibinfo {author} {\bibfnamefont {M.}~\bibnamefont
  {Guzzo}}, \bibinfo {author} {\bibfnamefont {J.~J.}\ \bibnamefont {Kas}},
  \bibinfo {author} {\bibfnamefont {L.}~\bibnamefont {Sponza}}, \bibinfo
  {author} {\bibfnamefont {C.}~\bibnamefont {Giorgetti}}, \bibinfo {author}
  {\bibfnamefont {F.}~\bibnamefont {Sottile}}, \bibinfo {author} {\bibfnamefont
  {D.}~\bibnamefont {Pierucci}}, \bibinfo {author} {\bibfnamefont {M.~G.}\
  \bibnamefont {Silly}}, \bibinfo {author} {\bibfnamefont {F.}~\bibnamefont
  {Sirotti}}, \bibinfo {author} {\bibfnamefont {J.~J.}\ \bibnamefont {Rehr}}, \
  and\ \bibinfo {author} {\bibfnamefont {L.}~\bibnamefont {Reining}},\
  }\href@noop {} {\bibfield  {journal} {\bibinfo  {journal} {Phys. Rev. B}\
  }\textbf {\bibinfo {volume} {89}},\ \bibinfo {pages} {085425} (\bibinfo
  {year} {2014})}\BibitemShut {NoStop}%
\bibitem [{\citenamefont {Onsager}\ \emph {et~al.}(1966)\citenamefont
  {Onsager}, \citenamefont {Mittag},\ and\ \citenamefont
  {Stephen}}]{onsager_integrals_1966}%
  \BibitemOpen
  \bibfield  {author} {\bibinfo {author} {\bibfnamefont {L.}~\bibnamefont
  {Onsager}}, \bibinfo {author} {\bibfnamefont {L.}~\bibnamefont {Mittag}}, \
  and\ \bibinfo {author} {\bibfnamefont {M.~J.}\ \bibnamefont {Stephen}},\
  }\href@noop {} {\bibfield  {journal} {\bibinfo  {journal} {Ann. Phys.}\
  }\textbf {\bibinfo {volume} {473}},\ \bibinfo {pages} {71} (\bibinfo {year}
  {1966})}\BibitemShut {NoStop}%
\bibitem [{\citenamefont {Ziesche}(2007)}]{ziesche_self-energy_2007}%
  \BibitemOpen
  \bibfield  {author} {\bibinfo {author} {\bibfnamefont {P.}~\bibnamefont
  {Ziesche}},\ }\href@noop {} {\bibfield  {journal} {\bibinfo  {journal} {Ann.
  Phys.}\ }\textbf {\bibinfo {volume} {16}},\ \bibinfo {pages} {45} (\bibinfo
  {year} {2007})}\BibitemShut {NoStop}%
\bibitem [{\citenamefont {Glasser}\ and\ \citenamefont
  {Lamb}(2007)}]{glasser_analysis_2007}%
  \BibitemOpen
  \bibfield  {author} {\bibinfo {author} {\bibfnamefont {M.~L.}\ \bibnamefont
  {Glasser}}\ and\ \bibinfo {author} {\bibfnamefont {G.}~\bibnamefont {Lamb}},\
  }\href@noop {} {\bibfield  {journal} {\bibinfo  {journal} {J. Phys. A}\
  }\textbf {\bibinfo {volume} {40}},\ \bibinfo {pages} {1215} (\bibinfo {year}
  {2007})}\BibitemShut {NoStop}%
\bibitem [{\citenamefont {Rojas}\ \emph {et~al.}(1995)\citenamefont {Rojas},
  \citenamefont {Godby},\ and\ \citenamefont {Needs}}]{rojas_space-time_1995}%
  \BibitemOpen
  \bibfield  {author} {\bibinfo {author} {\bibfnamefont {H.~N.}\ \bibnamefont
  {Rojas}}, \bibinfo {author} {\bibfnamefont {R.~W.}\ \bibnamefont {Godby}}, \
  and\ \bibinfo {author} {\bibfnamefont {R.~J.}\ \bibnamefont {Needs}},\
  }\href@noop {} {\bibfield  {journal} {\bibinfo  {journal} {Phys. Rev. Lett.}\
  }\textbf {\bibinfo {volume} {74}},\ \bibinfo {pages} {1827} (\bibinfo {year}
  {1995})}\BibitemShut {NoStop}%
\end{thebibliography}
\end{document}